\documentstyle[12pt]{article}
\begin{document}
\vspace*{2cm}
\begin{center}
{\Large\bf QUANTUM EINSTEIN GRAVITY AS A TOPOLOGICAL FIELD
THEORY}\\[1.5cm]
{\large Andrew Toon\footnote{Supported by the Royal Society England.}\\
University of Oxford\\
Department of Theoretical Physics\\
1 Keble Road\\
Oxford, OX1 3NP\\
England.}
\end{center}
\vspace*{1in}
\begin{abstract}
General covariance in quantum gravity is seen once one integrates over
all possible metrics. In recent years topological field theories have
given us a different route to general covariance without integrating
over all possible metrics. Here we argue that Einstein quantum
gravity may be viewed of as a topological field theory as long as a
certain constraint from the path integral measure is satisfied.
\end{abstract}
\newpage
\section{Introduction}
Topological quantum field theories first made their impact by the way
they reproduce topological invariants of certain manifolds [1,2]. From a
physical point of view it is hoped that these theories may have something
to do with quantum gravity. That is, one imagines that the initial
phase of the Universe is in an unbroken phase where there is no
space-time metric. The Universe we see today would then correspond to
the so called broken phase where, via some mechanism, the initial
unbroken phase developes some metric dependence there by generating a
space-time structure where physics can operate.

Something along these lines is witnessed in 2+1 dimensions where 2+1
dimensional quantum gravity can be given a gauge theoretic
interpretation in terms of a Chern-Simons theory with gauge group
$ISO(2,1)$, $SO(3,1)$ or $SO(2,2)$, depending on whether the
cosmological constant $\lambda$ is zero, positive or negative [3,4].
This theory has two natural phases depending on wheather a certain
operator has zero modes or not and it is argued that each of
these phases corresponds to the unbroken and broken phases of gravity [4,5].

Unfortunatly, if one tries to give a similar analysis to 3+1
dimensional quantum gravity, one soon fails for the simple reason that
quantum gravity in 3+1 dimensions has no natural gauge
theoretic interpretation like its 2+1 dimensional counterpart. There do exist
various 3+1 dimensional topological field theories but at present they
seem to have nothing to do with quantum gravity [6].

Rather then to try and make contact with gravity starting from a
topological field theory we try here to formally start with quantum
Einstein gravity and attempt to make contact with topological field theory.
That is, we argue that Einstein quantum
gravity can formally be given
an interpretation as a topological field theory\footnote{Strictly
speaking a topological field theory is only an approximation to a
quantum theory of gravity since we expect to summ over all possible
topologies.} .

This should come as no great suprise since in the path integral
approach to quantum gravity one integrates over all possible metrics,
in a given topology, and so one expects to extract only topological
information. But exactly what topology does one integrate all possible
metrics over and how does one formally do this in a quantum theory of gravity?

To answer this question we turn first of all to topological field
theories. Here we learn what needs to be done for the case of quantum
gravity in order that the theory as an interpretation as a topological
field theory. The crucial observation being the role played by the so
called gauge fixing metric. We then turn to the gauge fixing procedure
of Einstein
quantum gravity and argue that this can be given a formal structure
that mimics the proof of topological invariance in topological quantum
field theories based on BRST invariance. We then worry about the path
integral measure which must be independent of the gauge fixing metric
for the proof of topological invariance to be legal. From this we
derive a constraint on the particle content of a theory for
topological invariance to be preserved in the quantum theory.
We finally finish with a conclusion.

\section{Topological Field Theories}
In this section we will review the relevent facts of topological field
theories in order to discover what is needed in the case of quantum
gravity for it to have an interpretation as a topological field
theory. As we will see, the crucial point is in the gauge fixing. It
is this procedure that really tells us what we mean by a topological
field theory.

Consider then an $n$-dimensional smooth manifold $M_{n}$. We assume
that we can construct some classical action $S$ on $M_{n}$ which is
general coordinate invariant. That is, $S$ does not contain any
space-time metric of $M_{n}$. To evaluate, for example, the partition
function of $S$ we must gauge fix. The important point here is in order
to fully gauge fix the theory we must choose some metric $g_{ij}$ on
$M_{n}$. The fully gauged fixed quantum action $S_{q}$ then takes the
form\footnote{We will ignore such issues as Gribov ambiguities in this
paper.}:
\begin{equation}
S_{q}[\Phi_{r}, g_{ij}]=S[\Phi_{r}]+\delta_{Q}V[\Phi_{r},g_{ij}].
\end{equation}
Here, $\Phi_{r}$ $(r=1,2,...)$ are the fields of the theory including
matter, gauge, ghost and auxiliary fields etc. The second term on the
right hand side of equation (1) is the ghost plus gauge fixing term
associated with all the gauge invariance of $S[\Phi_{r}]$. $\delta_{Q}$
stands for the related nilpotent BRS-transformation and thus the
entire gauge fixing plus ghost term is written as a BRST variation of
some functional $V[\Phi_{r}, g_{ij}]$. The partition function for the
theory is thus given by:
\begin{equation}
Z(M_{n},g_{ij})=\int D[\Phi]\exp iS_{q}[\Phi_{r}, g_{ij}].
\end{equation}
What we mean by a topological field theory is that, for example, the
partition function only depends on the gauge fixing metric
$g_{ij}$ topologically. That is:
$$\frac{\delta}{\delta g^{ij}}Z(M_{n},g_{ij})=\int
D[\Phi]\exp(iS_{q})\frac{\delta}{\delta g^{ij}}(\delta_{Q}V)$$
\begin{equation}
=\int D[\Phi]\exp(iS_{q})\delta_{Q}(\frac{\delta}{\delta g^{ij}}V)=0,
\end{equation}
by BRST invariance. What is also very important is the assumption that
the path integral measure $D[\Phi]$ is metric independent. We will
return to this point shortly.

Our main point in this section has been the following. By a
topological field theory we mean that the partition
function, for example, of the theory does not depend explicitly on
the gauge fixing
metric $g_{ij}$ that appears only in the gauge fixing and ghost terms
of the quantum action. It depends on it only
topologically in that a small variation of the metric $g_{ij}$ will
not change $Z(M_{n},g_{ij})$. It is this aspect of topological field
theories that we claim to observe in Einstein quantum gravity.
\section{Gauge Fixing in Quantum Gravity}
Einstein quantum gravity is a non-renormalisable theory at least
pertubatively. It may, however, be a sensible theory in the
non-perturbative regime. With this view in mind, we wish to argue that
the full non-perturbative partition function of Einstein quantum
gravity has a structure similar to that of the previous section. That is,
Einstein quantum gravity has an interpretation as a topological field
theory with respect to some gauge fixing metric $g_{ij}$.

Our starting point is the usual classical action for Einstein
gravity in $n$ space-time dimensions:
\begin{equation}
S=\int\sqrt{G}(R+\lambda),
\end{equation}
where $R$ is the scalar curvature, $\lambda$ is the cosmological
constant and $G=detG_{ij}$ with $G_{ij}$ being the usual Einstein
metric of general relativity. Due to the invariance of $S$ under
diffeomorphims given by:
\begin{equation}
\delta_{f}
G_{ij}=G_{ik}\partial_{j}f^{k}+G_{kj}\partial_{i}f^{k}+(\partial_{k}G_{ij})
f^{k},
\end{equation}
$f^{i}$ being an arbitrary infinitesimal contravariant vector,
we must gauge fix equation (4) in order to quantise the theory. In
order to mimic the corresponding case in topological field theories we
first make the following observation.

We regard the metric $G_{ij}$ of equation (4) to be a rank
two symmetric tensor field on some manifold with metric $g_{ij}$ which we
regard as being a gauge fixing metric. Since the
classical action (4) is independent of $g_{ij}$, which will only appear
in the gauge fixing terms, we can trivially regard
the classical action (4) to be independent of $g_{ij}$.

For the gauge fixing condition $F_{i}$, we will choose the
familiar one given by $F_{i}=D^{j}G_{ij}+\frac{1}{2}\alpha b_{i}$
where $b_{i}$ is a Lagrange multiplier field, $D^{j}$ is the
gravitational covariant derivative with respect to some reference
(gauge fixing) metric $g_{ij}$ and $\alpha$ is a gauge fixing
parameter.
It is now
clear that the full action of the gauge fixing plus FP ghosts is given
by [7,8]:
\begin{equation}
S_{gf+FP}=\delta_{Q}\int\sqrt{g}(\bar{c}^{i}(D^{j}G_{ij}+\frac{1}{2}\alpha
b_{i})),
\end{equation}
where $\delta_{Q}$ is the associated BRS-transformation given by:
$$\delta_{Q}G_{ij}=c^{k}\partial_{k}G_{ij}+G_{ij}\partial_{k}c^{k}+G_{kj}
\partial_{i}c^{k},$$
$$\delta_{Q}c^{i}=c^{k}\partial_{k}c^{i},$$
\begin{equation}
\end{equation}
$$\delta_{Q}\bar{c}^{i}=b^{i},$$
$$\delta_{Q}b^{i}=0.$$
One may also check that the BRST operator $\delta_{Q}$ is nilpotent.
That is:
\begin{equation}
\delta_{Q}^{2}=0,
\end{equation}
for all fields.
Following the arguments of section (2), it is now clear that the
formal partition function of Einstein quantum gravity is topological
with respect to the gauge fixing metric $g_{ij}$. That is, given that:
\begin{equation}
Z(g_{ij})=\int D[G_{ij}]D[b^{i}]D[c^{i}]D[\bar{c}^{i}]\exp i(S+S_{gf+FP}),
\end{equation}
and assuming
that the path integral measure is independent of $g_{ij}$, then:
\begin{equation}
\frac{\delta}{\delta g_{ij}}Z(g_{ij})=0.
\end{equation}
Just like the case of section (2) regarding topological field
theories, we see that the formal partition function of pure Einstein
quantum gravity has exactly the same structure with respect to some
gauge fixing metric $g_{ij}$.
\section{The path integral measure}
We showed in the previous section that the formal partition function
of Einstein quantum gravity does not depend on a choice of gauge
fixing metric $g_{ij}$. In other words, Einstein quantum gravity is
topological with respect to $g_{ij}$. What was crucial in the above
observation is the assumption that the path integral measure of the
theory be independent of $g_{ij}$. This, as we shall now discuss, is
in general not true.

Consider any general coordinate invariant field theory in
$n$-dimensions with quantum action given by equation (1).
The partition function is then not the naive
partition function of equation (2) but is given by:
\begin{equation}
\tilde{Z}(M_{n},g_{ij})=\int\tilde{D}[\Phi]\exp
iS_{q}[\Phi_{r},g_{ij}],
\end{equation}
where $\tilde{D}[\Phi]$ stands for an appropriate general coordinate
invariant measure over the full set of fields $\Phi_{r}$. Fujikawa
[9,10]
as proposed the following choice of the measure for being general
coordinate invariant:
\begin{equation}
\tilde{D}[\Phi]=\prod_{x}[\prod_{r}d\tilde{\Phi}_{r}(x)],
\end{equation}
where for any given field component $\Phi_{r}$:
\begin{equation}
\tilde{\Phi}_{r}(x)=g^{\alpha_{r}}(x)\Phi_{r}(x),\; g(x)=\det g_{ij}.
\end{equation}
The $\alpha_{r}$ are constants which depend upon the tensor nature of
the field as well as the number of space-time dimensions. Some
examples being:
\begin{equation}
\alpha_{r}=\left\{ \begin{array}{ll}
                  \frac{1}{4} & \mbox{for each scalar field}\\
  \frac{n-2}{4n} & \mbox{for each component of a covariant vector
field}\\
\frac{n+2}{4n} & \mbox{for each component of a contravariant vector}\\
\frac{n-4}{4n} & \mbox{for each component of a covariant tensor field of rank
two.}
\end{array}
\right.
\end{equation}
It is now clear that:
\begin{equation}
\prod_{x}d\tilde{\Phi}_{r}(x)=\prod_{x}d[g^{\alpha_{r}}(x)\Phi_{r}(x)]=
\prod_{x}[g^{\alpha_{r}\sigma_{r}}(x)d\Phi_{r}(x)],
\end{equation}
where the signature $\sigma_{r}$ is +1 (-1) for commuting
(anti-commuting) fields. Thus, the general covariant Fujikawa measure
becomes:
\begin{equation}
\tilde{D}[\Phi]=\prod_{x}[g(x)]^{K}(\prod_{r,y}d\Phi_{r}(y)),
\end{equation}
where:
\begin{equation}
K=\sum_{r}\sigma_{r}\alpha_{r},
\end{equation}
is an index which measures the $g$-metric dependence of the path
integral measure. The partition function (11) thus becomes:
\begin{equation}
\tilde{Z}(M_{n})=(\prod_{x}[g(x)]^{K})Z(M_{n}),
\end{equation}
where $Z(M_{n})$ is the partition function using the naive measure
$\prod_{x,r}[d\Phi_{r}(x)]$. Thus for the topological invariance to be
preserved  at the quantum level with respect to the metric $g$, we
require:
\begin{equation}
K=\sum_{r}\sigma_{r}\alpha_{r}=0.
\end{equation}
\section{Path integral measure and quantum gravity}
In section (3) we argued that Einstein quantum gravity has an
interpretation as a topological field theory with respect to some
gauge fixing metric $g_{ij}$. What was crucial in this observation was
the asumption that the path integral measure be independent of the
gauge fixing metric $g_{ij}$. The last section, however, showed that
for this to be true, we require the index $K$ to vanish. For pure
Einstein quantum gravity in $n$-dimensions this is not the case.

Looking back at equation (9), we see that the full set of dynamical
fields that appear in the formal expression for the partition function
of Einstein quantum gravity is:
\begin{equation}
\{G_{ij}, b^{i}, c^{i}, \bar{c}^{i}\}.
\end{equation}
Here $G_{ij}$ is the metric that appears in the Einstein action which
we are regarding as a rank two symmetric tensor field on a manifold
with metric $g_{ij}$, $b^{i}$ is a Lagrange multiplier field which
enforces the gauge fixing constraint and $c^{i}$, $\bar{c}^{i}$ are the
corresponding ghost, anti-ghost fields. For this set of fields we find
that $K$ is given by:
\begin{equation}
K=\frac{1}{8}(n^{2}-5n-8),
\end{equation}
and we remind the reader that $G_{ij}$ has $n(n+1)/2$ components and a
vector field has $n$ components in $n$ space-time dimensions.
It is easy to check that $K$ does not
equal zero for any integer value of $n$. Thus, it appears that
Einstein quantum gravity in $n$ dimensions does not have an
interpretation as a topological field theory in any space-time
dimension\footnote{We stress here that we are talking about Einstein
quantum gravity in the second order formalisim with metric $G_{ij}$
and not the first order formalism interms of the vierbein and spin
connection fields.}.

The above observation may be a blessing in disguise since many feel
that a true quantum theory of gravity should be a theory of everything
in that it should also tell us about the matter and field  content of
our Universe. Our idea then is to add some other fields to pure
Einstein gravity, without spoiling the quantum topological invariance
arguments of section (3), such that the index $K$ is forced to be zero [11].

Lets us
specialise to the physically interesting case of $n=4$. If we couple to
Einstein gravity a pure gauge theory with gauge group $\cal G$ our
classical
action  becomes:
\begin{equation}
S=\int\sqrt{G}((R+\lambda)+F^{ij}F_{ij}),
\end{equation}
where $F^{ij}$ is the assiociated field strength of the gauge field
$A_{i}$.
After introducing the appropriate Lagrange multiplier and ghost fields
for the gauge fixing of $A_{i}$ with respect to the gauge group $\cal G$,
the formal partition function for this
theory is now:
\begin{equation}
Z=\int D[G_{ij}]...D[A_{i}]...\exp iS_{q},
\end{equation}
where the dots after the $G_{ij}$ measure represent the fields
needed in the gauge
fixing of diffiomorphism invariance and the dots after the $A_{i}$
measure represent
the fields needed in the gauge fixing of the gauge field
$A_{i}$ with respect to the gauge group $\cal G$. Recall also that
the gauge fixing of the gauge field $A_{i}$, as well as the inclusion
of its kinetic term,  is performed with respect
to the Einstein metric $G_{ij}$ and thus the proof of topological
invariance with respect to the gauge fixing metric $g_{ij}$ of
section (3) is maintained.

We see that $K$ now takes the value:
\begin{equation}
K=-\frac{6}{4}+\frac{dim({\cal G})}{4},
\end{equation}
where $dim({\cal G})$ is the dimension of the gauge group $\cal G$, in this
case, the dimension of the adjoint representation of $\cal G$. We clearly see
that for $K$ to be zero $dim({\cal G})=6$. That is, if we couple a pure gauge
theory with $dim({\cal G})=6$ to Einstein gravity, the full quantum theory has
an
interpretation as a topological field theory with respect to the gauge
fixing metric $g_{ij}$.

There are a number of choices of gauge group $\cal G$ such that
$dim({\cal G})=6$.
An obvious choice is the product of six $U(1)$'s but a more
interesting one is ${\cal G}=SO(4)\sim SU(2)\times SU(2)$. We thus see, in
particular, that if we couple a pure $SO(4)$ gauge theory to Einstein
gravity in $n=4$ space-time dimensions then $K$ is forced to be zero
and thus the full quantum theory has an interpretation as a
topological field theory with respect to the gauge fixing metric
$g_{ij}$.

This is an interesting observation in that an $SO(4)$
gauge theory is essentially a spin connection field in $n=4$
dimensions with Euclidean metrics. Perhaps the theory is telling us
that a proper approach to quantum gravity is via the vierbein-spin
connection formalism which has been so succesful in 2+1 dimensions.
\section{conclusion}
In this paper an attempt as been made to make a connection between
topological field theory and Einstein quantum gravity. Our approach
has been to start off with Einstein quantum gravity and argue that this
could be given an interpretation as a topological field theory with
respect to some gauge fixing metric $g_{ij}$. This being exactly what
is meant by topological invariance in topological field theories.
After worrying about the path integral measure it was discovered that
the path integral measure of pure Einstein quantum gravity in
$n$-dimensions develops a dependence on the gauge fixing metric
$g_{ij}$. This ruining the claim that pure Einstein quantum gravity has
an interpretation as a topological field theory with respect to the
gauge fixing metric $g_{ij}$. However, it was shown that if we couple
a $SO(4)$ pure gauge theory, for example, to Einstein gravity in $n=4$
space-time
dimensions then the corresponding
quantum theory has a path integral measure independent of the gauge
fixing metric and thus has an interpretation as a topological field
theory with respect to the gauge fixing metric $g_{ij}$.

This is, we believe, a very interesting observation. The main physical
motivation for studying topological field theories is the belief that
they have something to do with quantum gravity. In this paper we have
argued that Einstein gravity, in its second order formalism, indeed as
this interesting interpretation when coupled to other fields. If we
again return to a $n$-dimensional space time and couple Einstein
gravity to some gauge theory with gauge group $\cal G$ together with $m$
scalar fields transforming under some representation of $\cal G$, then the
general constraint that the corresponding path integral measure be
independent of the gauge fixing metric $g_{ij}$ becomes:
\begin{equation}
K=\frac{1}{8}(n^{2}-5n-8)+dim_{adj}({\cal
G})(\frac{n-3}{4})+\sum_{i=1}^{m}\frac{dim{\cal G}_{i}}{4}=0,
\end{equation}
where each scalar field transforms in some representation ${\cal G}_{i}$ under
$\cal G$. This equation is a remarkable equation in that in a single
constraint we see the number of space-time dimensions $n$, the
dimension of the adjoint representation $dim_{adj}{\cal G}$ of the gauge
group $\cal G$ together with the possible dimensions of the representations
${\cal G}_{i}$ the scalar fields can transform under. Since there are group
theoretic constraints between $\cal G$, $dim_{adj}({\cal G})$ and
$dim({\cal G}_{i})$,
one can imagine some classification of possible theories based on
equation (25) such that the corresponding theory has an interpretation
as a topological field theory.

Finally we would like to make a comment on including fermionic matter in
the theory. This of course requires a theory of gravity in its first
order formalism. That is, interms of the vierbein and spin connection
fields. In the quantum region,  this theory will be quite different from
the corresponding pure metric theory but we believe that such a theory
can also be given a topological interpretation as in 2+1 dimensions.
Indeed, the above
observations suggest, at least in $n=4$ space-time dimensions, the
inclusion of a spin connection field. This will be discussed
elsewhere.
\newpage
\begin{center}
{\large \bf References}
\end{center}

[1] E. Witten, Commun. Math. Phys. 117 (1988) 353.\\

[2] D. Birmingham, M. Blau, M. Rakowski, G. Thompson, Phys. Rep. 209\\

(1991) 129.\\

[3] E. Witten, Nucl. Phys. B311 (1988/89) 46.\\

[4] E. Witten, Nucl. Phys. B323 (1989) 113.\\

[5] A. Toon, Phys. Rev. D47 (1993) 2435. ^^ ^^ Background fields in
2+1 topological\\

 gravity'', to appear in Phys. Rev. D.\\

[6] G. Horowitz, Commun. Math. Phys. 125 (1989) 417.\\

[7] S. Adler, Rev. Mod. Phys. Vol. 54, No. 3 (1982) 729.\\

[8] Z. Bern, S. Blau, E. Mottola, Phys. Rev. D43 (1991) 1212.\\

[9] K. Fujikawa, Phys. Rev. Lett. 42 (1979) 1195.\\

[10] K. Fujikawa, O. Yasuda, Nucl. Phys. B245 (1984) 436.\\

[11] A. Toon ^^ ^^ Particle content in topological field theories'',
University of Oxford\\

 preprint: OUTP-92-30P.

\end{document}